\documentclass{svproc}
\usepackage{graphicx}
\usepackage{multirow}
\usepackage{ulem}
\usepackage{subfig}

\usepackage{url}

\begin{document}
\mainmatter              
\title{Cost of Dietary Data Acquisition with Smart Group Catering}
\titlerunning{Cost of Dietary Data}  
%
\author{Jiapeng Dong \and  Pengju Wang\and Weiqiang Sun}
\authorrunning{Jiapeng Dong et al.} 

%

\institute{Shanghai Institute for Advanced Communication and Data Science, Shanghai JiaoTong University, Shanghai, 200240, China \\
\email{sunwq@sjtu.edu.cn}}

\maketitle              

\begin{abstract}

The need for dietary data management is growing with public awareness of food intakes. As a result, there are increasing deployments of smart canteens where dietary data is collected through either Radio Frequency Identification (RFID) or Computer Vision(CV)-based solutions. As human labor is involved in both cases, manpower allocation is critical to data quality. Where manpower requirements are underestimated, data quality is compromised.
This paper has studied the relation between the quality of dietary data and the manpower invested, using numerical simulations based on real data collected from multiple smart canteens.
We found that in both RFID and CV-based systems, the long-term cost of dietary data acquisition is dominated by manpower.
Our study provides a comprehensive understanding of the cost composition for dietary data acquisition and useful insights toward future cost effective systems.

\keywords{ CV systems, data accuracy, dietary data acquisition, dietary management, health management, RFID systems, Smart group catering }
\end{abstract}

\section{Introduction}


The emerging public concern with health has led to the proliferation of health management applications, individual health monitoring\cite{peom:kyon} and nutritional assessments\cite{mark:ya,pari:sher} in which dietary data is often an important component.
Given the diversity of food and unpredictable dining locations, however, recording a person몶s regular intake has never been an easy task.
Since 2017, smart group catering (SGC) systems targeted at canteens of different sizes, and featuring automatic billing and data acquisition, have become popular.
Onsite experiences with SGC systems indicate that even though dietary data acquisition technologies seem to be readily available, data quality may vary drastically from one canteen to another. One important reason behind this is a widespread under-estimation of the necessary manpower needed for accurate data acquisition.
Thus, it is important to understand this cost and its relationship to other factors.



\begin{figure}[htbp]
\centerline{\includegraphics[width=0.8\textwidth]{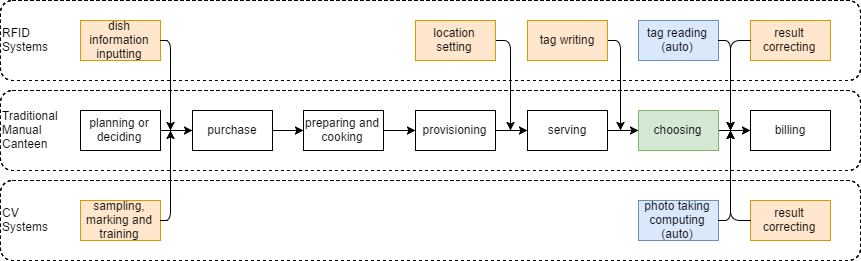}}
\caption{Workflow of Traditional Canteen and Intrusion of Two Systems}
\label{fig1}
\end{figure}

There are two types of widely used SGC systems implemented for dietary data acquisition몱Radio Frequency Identification (RFID)- and Computer Vision (CV)-based solutions. The data acquisition workflows of the two types of solutions are shown in Fig.~\ref{fig1}. In RFID-based systems, special dishes with embedded tags are used when food is served. The food information is read when customers checkout. In CV-based systems, cameras are used at checkout counters to recognize the dish . Fig.~\ref{fig1} shows the basic workflow of a traditional canteen and the extra procedural intrusion of the two systems.

RFID is a mature technology and has countless applications, \cite{ganj:jong} e.g. RFID systems were first proposed for SGC by Yao.X et al.\cite{yao:jian} in 2011. There were subsequent implementations by Y.H.Liang et al.\cite{lian:chen}, Pai-Hsun.C et al.\cite{pai:ying} and E.B.Kossonon et al.\cite{koss:wang}, which were mainly prototypes and somewhat different from the systems deployed in canteens, as shown in Fig.~\ref{fig1}. With the advantage of mature technology and simple software, RFID systems currently account for over 80\% of the market.

CV systems are less mature. The food detection algorithm used in CV systems has become popular in recent years. Lead by Bossard.L et al.\cite{boss:guil} in 2014, some researchers put up new datasets\cite{qian:jing}, while others focused on a food recognition algorithm. A group of studies tried to solve image segmentation before further recognition\cite{guo:dong,sine:gian}. Those that directly tackle the entire food detection mission\cite{marc:peti,yunan} also achieve good performance. Among all these studies, the sequential works\cite{gian:paol,edua:beat} are the milestone for the canteen scenarios, which also provides solid references for our study.
%
%

In this paper, we aimed to study the relation between dietary data quality and invested manpower. We conducted our study by means of numerical simulations, with parameters taken from real-life canteens operating SGC systems. Our contributions are:
\begin{itemize}
\item Information flow-centric modeling of dietary data acquisition for RFID and CV-based SGC systems.

\item A comprehensive numerical analysis of the cost of dietary data acquisition with the two types of technologies.

\item Comparative analyses of RFID and CV-based systems' application scenarios and limitations based on the dynamic relationships between cost, data accuracy and other relevant factors of deployment.


\item Future directions and evolutionary trends for dietary data collection using SGC systems.
\end{itemize}

The rest of the paper is organized as follows: Chapter~\ref{model} builds the cost accuracy model based on information flows and key procedure properties. Chapter~\ref{experiment} describes our data set and basic settings of the experiments. Chapter~\ref{result} progressively analyzes the cost of dietary data harvesting and its major influence factors. Other accessory factors are discussed in  Chapter~\ref{discussion}. Finally, the conclusions and future outlooks are presented.

\section{Models}\label{model}

In this section, we modeled different types of costs for both systems and the relationships between relevant ones and data accuracy. We included the cost of the system itself and the corresponding extra costs necessary to maintain normal operation and obtain accurate data.

\subsection{Cost Composition: Key Procedures and Cost Groups}

\begin{figure}[htbp]
\centerline{\includegraphics[width=0.8\textwidth]{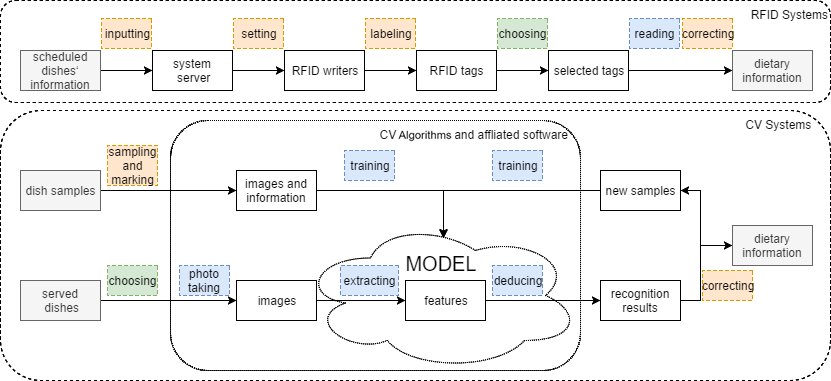}}
\caption{Information Flow of Two Systems: Step blocks in white background are the intermediate formation or carrier of dietary information and those in gray represent sources and targets.}
\label{fig2}
\end{figure}

In order to excavate the essential mechanism of dietary data harvesting in both RFID and CV systems, we reviewed its general workflow, as illustrated in Fig.~\ref{fig1}. We considered the nature of data harvesting as an information flow. Thus, we ruled out all the factors that were not concerned with the information flow. Afterwards, we defined and located the key procedures for the necessary transformation or transmission of dietary information. The rearranged and expanded workflow chart is presented as an information flow chart in Fig.~\ref{fig2}.

From Fig.~\ref{fig2}, the data quality of RFID systems is much more dependent on staff operation at three key sequential   procedures, namely inputting, setting and labeling. The flow of CV systems, however, is primarily determined by its algorithm performance and secondarily by sampling procedures. The information flow of RFID systems looks simpler and more direct while the flow of CV systems contains fewer procedures requiring staff operation. Both systems have to correct errors at checkout to ensure normal billing.

All costs are incurred during the procedure of information transformation and transmission. In order to discriminate between different cost items, we divided them into two groups according to whether they are staff operation related costs (SORC) or not staff operation related cost (NSORC). For each type of system:
\begin{equation}
SORC^{RFID}=C_{input}+C_{set}+C_{label}+C_{correct}\label{eq1}
\end{equation}
\begin{equation}
NSORC^{RFID}=\sum C^{RFID}_{devices}+C^{RFID}_{software}+C_{plate-loss}\label{eq2}
\end{equation}
\begin{equation}
SORC^{CV}=C_{sample}+C_{correct}\label{eq3}
\end{equation}
\begin{equation}
NSORC^{CV}=\sum C^{CV}_{devices}+C^{CV}_{software}\label{eq4}
\end{equation}
, where $C_{plate-loss}$ represents the RFID-embedded plates malfunctioning during the usage, determined by the plate number used per meal and the statistical loss rate. Assuming five years of lifespan, devices and their converted per meal cost for each system are presented in Table.~\ref{tab1}, where $m$ depends on the ratio of canteen total throughput and unit checkout velocity.

\begin{table}
\caption{NSORC Items Comprised in Two Systems}
\begin{center}
\begin{tabular}{r@{\quad}r@{\quad}r@{\quad}rl}
\hline
\multicolumn{1}{l}{\rule{0pt}{12pt}System}&\multicolumn{1}{l}{\rule{0pt}{12pt}NSORC Item}&\multicolumn{1}{l}{\rule{0pt}{12pt}Number}&
\multicolumn{2}{l}{Value (RMB)}\\[2pt]
\hline\rule{0pt}{12pt}
\multirow{8}*{RFID} & RFID Writer & $T$ & $1.37\times 10^{-1}$  &\\
& RFID Reader & $m$ & $5.48\times 10^{-1}$ & \\
& Control Terminal & $1$ & $5.48\times 10^{-1}$ & \\
& Checkout Terminal & $m$ & $5.48\times 10^{-1}$ & \\
& Peripheral Network & $1$ & $2.74\times 10^{-1}$ & \\
& Server & $1$ & $8.22\times 10^{-1}$ & \\
& RFID software & $1$ & $5.48$ & \\
& RFID plate & $n * rate$ & $3.5$ & \\
\hline
\multirow{4}*{CV} & Embedded Camera & $m$ & $8.22\times 10^{-2}$ & \\
& Checkout Terminal & $m$ & $8.22\times 10^{-1}$ & \\
& Server (extra training) & $1$ & $1.37$ & \\
& CV software & $1$ & $10.96$ & \\
\hline
\multicolumn{4}{l}{$^{\mathrm{a}}$$T$ for dish types and $n$ for total plate number.} & \\
\end{tabular}
\label{tab1}
\end{center}
\end{table}

\subsection{Factors of Accuracy: Staff Operation and Sample Accumulation}

From the information flow in Fig.~\ref{fig2}, we concluded that the data accuracy mechanism differed between the two systems. The RFID system contained three staff-operated procedures and did not generate any false data when all three procedures were operated without any failure. Although the CV system required sampling and marking procedures, the number of executions was much smaller compared to the scale of the data to be collected. Meanwhile, the data was determined by the deduction results, which made the performance of the CV system mainly dependent on the CV dish recognition model applied and the number of samples in the training set.


man-hours was adopted as the measure of staff work in different procedures. Moreover, we expanded the man-hour concept into equivalent man-hours (EMH) which was defined to equivalently measure the extra cost invested to harvest dietary data across distinct procedures. With regards to the staff's non-standard operation and corresponding accuracy, based on our on-site knowledge, the following assumptions were raised:
\begin{itemize}
\item The accuracy of a key procedure carried out by staff once was proportionate to the extra EMH that the staff was provided.
\item The accuracy of a key procedure always reached one hundred percent with sufficient extra EMH.
\item As the provided extra EMH increased, the marginal accuracy growth continuously decreased toward the endpoint of accuracy.
\item The pattern of the marginal accuracy growth variation differed per procedure according to their attributes.
\end{itemize}

Since the power function is the simplest function fitting all these assumptions, it was used to construct our EMH Accuracy (EMH-A) model:

\begin{equation}
Accuracy(h)=(\frac{h}{S})^{\alpha},\quad 0\leq \alpha\leq 1\label{eq5}
\end{equation}
, where S is the standard EMH needed when accuracy reaches one hundred percent, and alpha is the procedure distinction coefficient representing procedure features' effects on the marginal accuracy growth patterns. The specific $S$ values of inputting, setting, labeling and correction procedures are based on the average time taken in a real canteen environment. In addition, the knowledge and skills required by the procedures are also taken into account. The baseline values are listed in Table.\ref{tab2}

\begin{table}
\caption{Parameters of EMH Accuracy Model}
\begin{center}
\begin{tabular}{r@{\quad}r@{\quad}rl}
\hline
\multicolumn{1}{l}{\rule{0pt}{12pt}Procedure}&\multicolumn{1}{l}{\rule{0pt}{12pt}$S$ (hour)}&
\multicolumn{2}{l}{$\alpha$}\\[2pt]
\hline\rule{0pt}{12pt}
Inputting & $1.7 \times 10^{-1}$ &  $0.6$ & \\
Setting & $6.7 \times 10^{-2}$ &  $0.4$ & \\
Labeling & $1.39 \times 10^{-3}$ &  $0.1$ & \\
Correction & $1.1 \times 10^{-2}(1.1 \times 10^{-3})$ &  $0.15$ & \\
\hline
\end{tabular}
\label{tab2}
\end{center}
\end{table}

\begin{figure}[htbp]
\centerline{\includegraphics[width=0.40\textwidth]{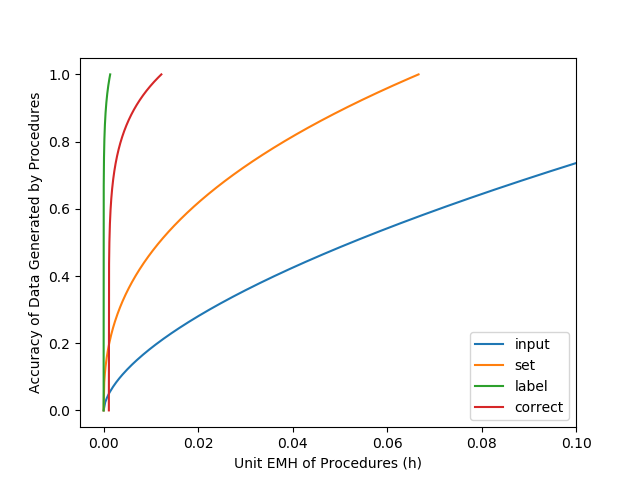}}
\caption{Accuracy by Unit EMH Cost of Four Procedures' EMH-A Models}
\label{fig3}
\end{figure}

The value of $\alpha$ depended on procedure features. The staff-related procedures in dietary data harvesting scenarios usually have three features: automation degree, throughput pressure, and internal complexity.
A procedure with more automation is more accurate, and thus the value should be bigger.
It is more difficult for a pressured procedure to reach high accuracy, which leads to a small value.
Higher accuracy can be easier to achieve with lower EMH if the procedure is comparatively simple, so the value should be small.
A binding system like RFID usually has high pressured labeling and correction procedures, a less pressured but more complex and automatic setting procedure and a more complex but less pressured inputting procedure. Therefore, using the values of $\alpha$ shown in Table.~\ref{tab2}, the curves of each procedure are drawn in Fig.~\ref{fig3}, where the EMH-axis offset of the correction procedure is determined by the fixed cost of total price correction, which is crucial for normal billing.

As in the previous discussion, the algorithm used in the CV model is the major internal factor for data accuracy. The procedure of information inputting including sample dish preparation, sampling (photo taking), and marking, with time reserved for training, was also taken into consideration.
The cost of this procedure is very high, about 0.33 h of EMH. This forces the on-site manager to minimize the sample number and to use the served dish for new samples at checkout as much as possible. Based on the general characteristics of the CV learning model, three assumptions were raised to model the relationship between deduction accuracy and sample size used in training:
\begin{itemize}
\item The performance of deduction always has a less-than-one upper bound decided by the algorithm the model applies.
\item When the accuracy is low, the increment caused by sample number increase is prominent.
\item As the accuracy approximates its upper bound, its marginal growth drops increasingly rapidly.
\end{itemize}

We chose the sigmoid function as our prototype to approximate the actual learning process of the dish recognition model. The Sample Number Accuracy (SNA) model is as follows:

\begin{equation}
Accuracy(n_{sample})=U * sigmoid(\beta * n_{sample})\label{eq6}
\end{equation}
, where $n_{sample}$ represents the number of samples, $U$ the upper bound, and $\beta$ is the transmission coefficient representing the algorithm's feature extraction efficiency. Based on the algorithm performance in \cite{edua:beat} and actual deployment situations, the value of $U$ was set to 0.85.


\section{Experiments}\label{experiment}

\subsection{Data Sets and Baseline Parameters}

Our data set was collected from thirteen canteens throughout mainland China. The types of canteens included government departments, primary schools, colleges, private and state enterprises. The data content contained the SGC system deployment profile of each canteen, menu update records of over four hundred dish types and over a million dish transaction details over a time span of more than half a year. The data set established a numerical foundation for our simulations.

According to previous procedure and information flow analyses, four main canteen features which have major impact on data acquisition cost were extracted. These four canteen parameters are listed in Table.~\ref{tab3}, with their definitions and units. The values here were based on the profile of our most familiar canteen, and were used as the baseline in the experiments. In addition, the product of $T$ and $N$ represents the canteen's scale, i.e., the customer capacity, while $F$ and $R$ show the canteen's service quality.

\begin{table}[h]
\caption{Canteen Feature Parameters}
\begin{center}
\begin{tabular}{r@{\quad}r@{\quad}r@{\quad}rl}
\hline
\multicolumn{1}{l}{\rule{0pt}{12pt}Parameter}&\multicolumn{1}{l}{\rule{0pt}{12pt}Definition}&\multicolumn{1}{l}{\rule{0pt}{12pt}Unit}&
\multicolumn{2}{l}{Value}\\[2pt]
\hline\rule{0pt}{12pt}
T & per meal dish types & types / meal & 20 & \\
N & dish number of each type & dishes / type & 70 & \\
F & frequency of adding new dish type & types / meal & 0.3 & \\
R & rotation of old dish type & types / meal & 6 & \\
\hline
\end{tabular}
\label{tab3}
\end{center}
\end{table}

\subsection{Basic System Characteristics and Experimental Settings}

Here we briefly look into the basic characteristics of the two systems in order to make preliminary settings for experiments.

\subsubsection{Cost Allocation of RFID Systems}

RFID systems comprise of three sequential key procedures among which cost can be allocated in various proportions. The function between a set of EMH costs, i.e., $(H_{input}, H_{set}, H_{label})$ and the corresponding accuracy, cannot be clearly depicted in graphs. After changing the $F$ of our canteen baseline to zero for simplification and better demonstration, we were able to draw the accuracy contour by the summed EMH cost of setting and labeling in Fig.~\ref{fig4}.


In this simplified condition, we can prove that for each convex accuracy contour, there is a point where the total cost of the two procedures is optimal. These points, as are shown in Fig.\ref{fig4}, form an optimal path joined first by the tangent points of the auxiliary total cost line with the contour, and then by the intersections of the line with the upper boundary of setting. The segmentation of optimal path like this is common since the cost of labeling is major in most conditions.

%
In summary, there is always an optimal path for EMH cost allocation and the path can be approximated as a broken line with several fixed proportions. Therefore, the path is adopted wherever relevant throughout this paper.

\subsubsection{Cold Start Problem of CV Systems}

\begin{figure}[!ht]
 \subfloat[RFID Systems: Accuracy Contour and Optimal Allocation Path\label{fig4}]{%
   \includegraphics[width=0.48\textwidth]{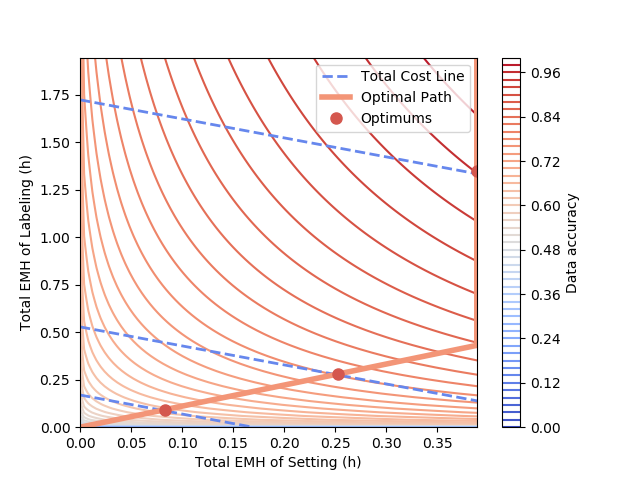}
 }
 \hfill
 \subfloat[CV Systems: Accuracy and Total EMH of First Hundred Meals\label{fig5}]{%
   \includegraphics[width=0.48\textwidth]{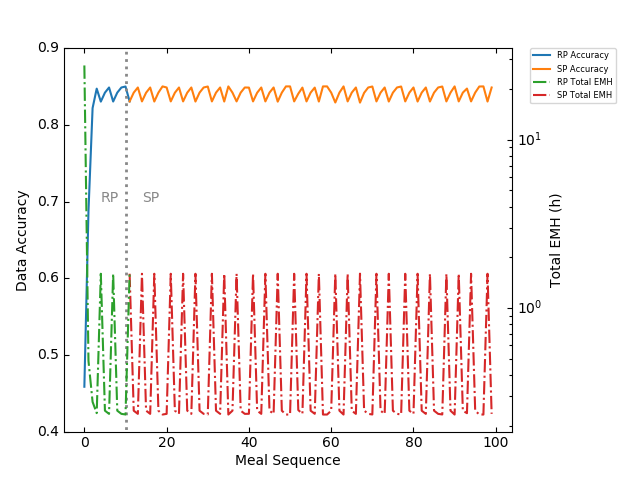}
 }
 \caption{Basic Characteristics}
 \label{fig40}
\end{figure}

CV systems are stateful and their accuracy depends on the size of the training set. Accuracy is attainable, but a ramping-up process is necessary considering the high cost of sampling which the canteen usually tends to skip or cut down. Thus, as demonstrated in Fig.~\ref{fig5}, the cold start problem of CV systems results in the ramping-up period (RP) of accuracy which zooms at the very beginning and ends when the differential accuracy increment (3-meal-long window) drops below $1\times 10^{-5}$. The rest part is defined as the stable period (SP) because the accuracy only fluctuates in a comparatively small range. The cost decreases sharply during the RP, especially between the first two meals (about a hundredth), and in SP peaks in accordance with the frequency of new dish addition, with an overall low average afterwards.

Extra EMH in RP, which can be multiple times that of SP, is essential and requires preparation in advance. Fortunately, the cold start problem only happens when a great many types of new dishes need to be sampled, and in most conditions only once. The infrequent occurrence of the problem and the large EMH gap between the two periods make it possible and reasonable to ignore the RP in a long term study. We will regard the EMH of the CV system to be specialized in SP unless otherwise mentioned.


\section{Results}\label{result}

In order to conduct a thorough study into the cost of dietary data harvesting with RFID and CV systems, we started our baseline from the general cost and its composition. Then we focused on the SORC dynamics of the two systems under variable accuracy targets. After another set of experiments conducted under various canteen conditions, four typical canteens
were specified and used as examples for straightforward understanding.

\subsection{General Cost Composition}

\begin{figure}[htbp]
\centerline{\includegraphics[width=0.48\textwidth]{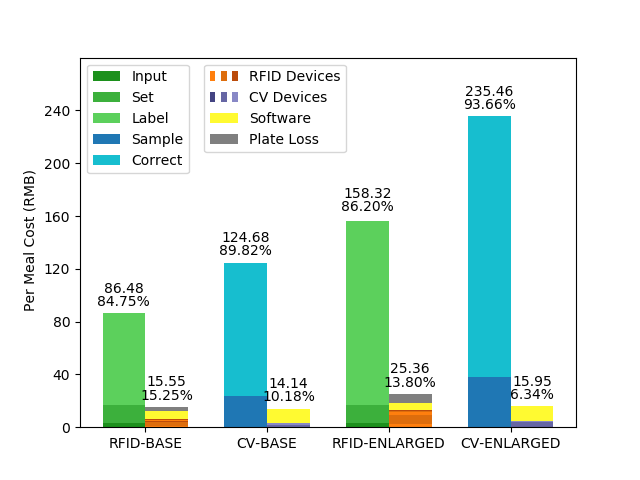}}
\caption{Mealy Cost of 100\% Target Accuracy by Cost Group, System and Canteen Scale: SORC(left) and NSORC(right) for each system and canteen.}
\label{fig60}
\end{figure}

Simulations were carried out with the two separate systems to evaluate the total cost of dietary data harvesting of a single meal with target accuracy of 1 in the baseline canteen. For a better comparison of NSORC in canteens of different scales, another set of experiments on an enlarged canteen were also appended. For experiment settings, NSORC items' statistical prices and baseline values of model parameters were adopted as listed in TABLE.~\ref{tab1}. The enlarged canteen was set with twice the customers (900 people) and more dish types (50 types). Furthermore, the values of EMH were transformed into RMB by current average hour wage level. In this way, we had four groups of total cost with composition as shown in Fig.~\ref{fig60}.

NSORC accounted for around 10\% to 15\%, much less than SORC does in both systems and both canteens. The sum of NSORC of both systems in the baseline canteen differs little (about 1\% of total), but as the scale of the canteen increases, the NSORC of RFID systems increases faster than CV systems, due to its extra plate loss (about 1.9\% of total) and RFID writing devices bonded to dish type numbers (about 2.2\% of total). Meantime, the impact on CV systems from device addition appears minimal with barely a 0.7\% increment of the total.

Compared to the insignificant increase of NSORC, SORC ascends synchronously with the canteen scale both in sum and ratio. Based on the invariance and minority role played by NSORC in data accuracy, SORC is more worthy of further study.

\subsection{SORC by Target Accuracy}\label{cv-need-crt}

In this section we explore the relationship between the SORC and data accuracy, in a specific baseline canteen scenario. As was illustrated above, RFID systems' accuracy could be assigned arbitrarily between 0 and 1 while that of CV systems was fixed at the average level of the stable period without manual data correction at checkouts. The higher accuracy of CV systems required manual data correction at checkout. The same correction could also be available for RFID systems but will be discussed later.

\begin{figure}[!ht]
\subfloat[Cost Composition\label{fig61}]{%
\includegraphics[width=0.48\textwidth]{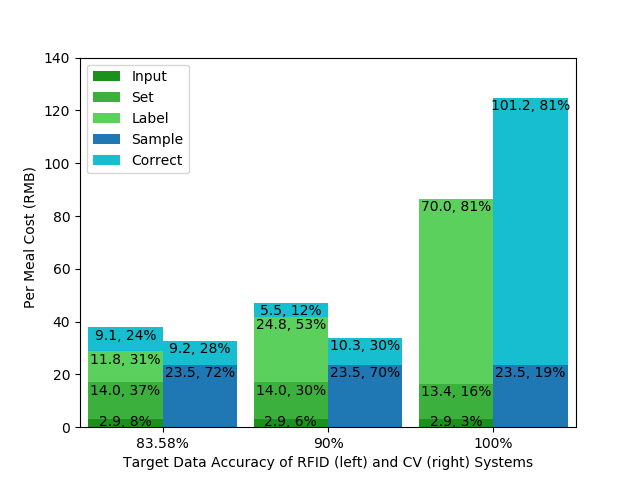}
}
\hfill
\subfloat[Total Cost Comparison\label{fig62}]{%
\includegraphics[width=0.48\textwidth]{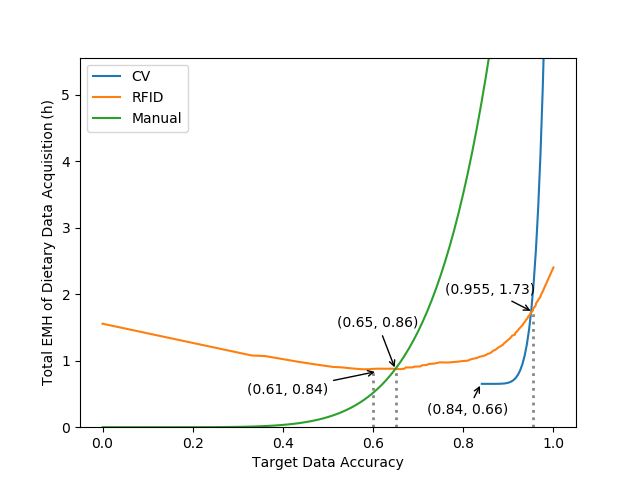}
}
\caption{SORC by Target Accuracy}
\label{fig:dummy}
\end{figure}

Experiments were performed conforming with the baseline canteen and with three accuracy targets, as is shown in Fig.~\ref{fig61}. The diagram describes the distinction of the two systems in general: With base SP accuracy, the cost of the labeling procedure accounts for about 33\% of RFID systems, while in CV systems the sampling procedure dominates. When the target accuracy rises, despite the correction cost decrement, the labeling procedure takes up almost the entire cost increment (about 25.7\% and 52.3\% of total) while setting and inputting procedures remain nearly the same. When it comes to CV systems, with the invariant sampling cost, the correction cost increment is minor (about 3.1\% of total) when the accuracy grows to 90\%, but it then increases to full accuracy. Therefore, the target accuracy was expanded to the whole range. The results of more detailed experiments are demonstrated in Fig.~\ref{fig62}. The cost curve when data is manually harvested is also provided for comparison.

%
In the baseline scenario, the RFID system out-performed the CV system only when accuracy was above 0.96. The curve of the RFID system in Fig.~\ref{fig62} showed convexity to some degree, with a minimal cost around accuracy of 0.6, because of the increased total price correction cost at low accuracy and the high cost required for high accuracy. Without data correction, the accuracy of CV systems stabilized at 0.84 and increased to 0.92. This efficiency was lost when the power law of the correction cost dominated and the cost surpassed the level of RFID systems (around 0.96).

To summarize, compared to the manual means of data harvesting at checkout, the SGC systems take the lead beginning from 0.63 accuracy by CV systems and then after 0.95 accuracy by RFID systems. Deploying an SGC system can save over 80\% of the cost of dietary data harvesting with an accuracy greater than 0.8.

%
%

\subsection{SORC by Canteen Features}

To extend our analysis to more canteens in real circumstances, experiments were designed to determine how the four features of canteens, i.e., $T$, $N$, $F$ and $R$ as listed in TABLE.~\ref{tab3}, influenced the accuracy of harvested data and the cost. $T$ and $N$ were grouped together as were $F$ and $R$. Since a dynamic analysis was required and the effect of target accuracy has already been studied, we concentrated on the average cost and accuracy of CV systems' stable period and the cost across five different target accuracies of RFID systems, that is, $0.6, 0.7, 0.8, 0.9, 1$.

\subsubsection{Dish Type Numbers and Dish Number of Each Type}

\begin{figure}[!ht]
\subfloat[by $T$ of RFID\label{fig70}]{%
\includegraphics[width=0.24\textwidth]{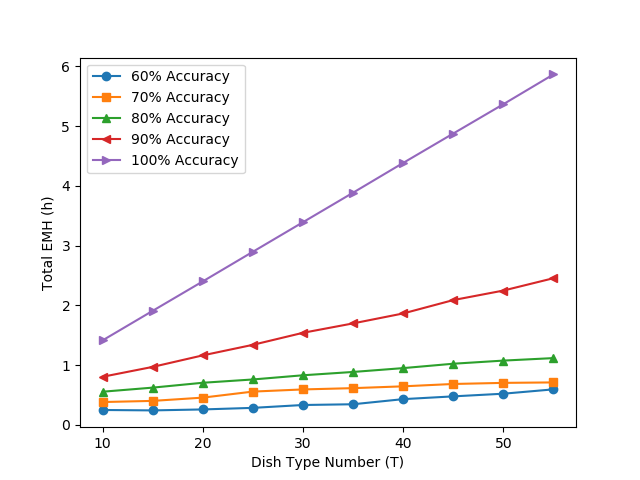}
}
\hfill
\subfloat[by $N$ of RFID\label{fig71}]{%
\includegraphics[width=0.24\textwidth]{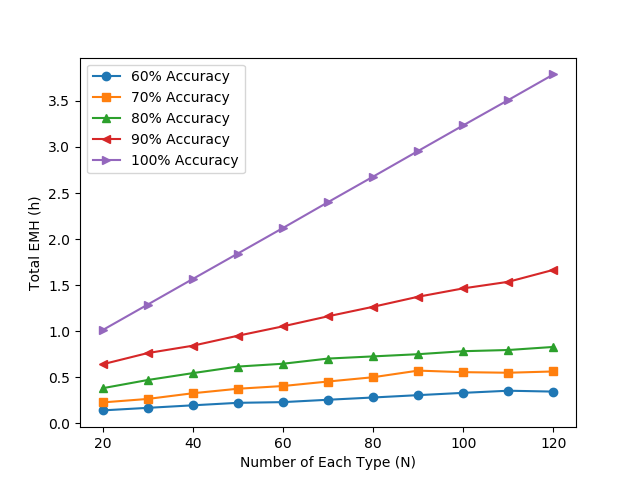}
}
\hfill
\subfloat[by $T$ of CV\label{fig72}]{%
\includegraphics[width=0.24\textwidth]{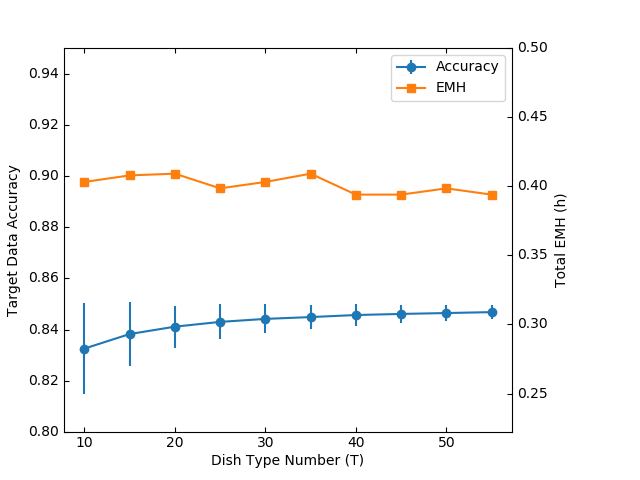}
}
\hfill
\subfloat[by $N$ of CV\label{fig73}]{%
\includegraphics[width=0.24\textwidth]{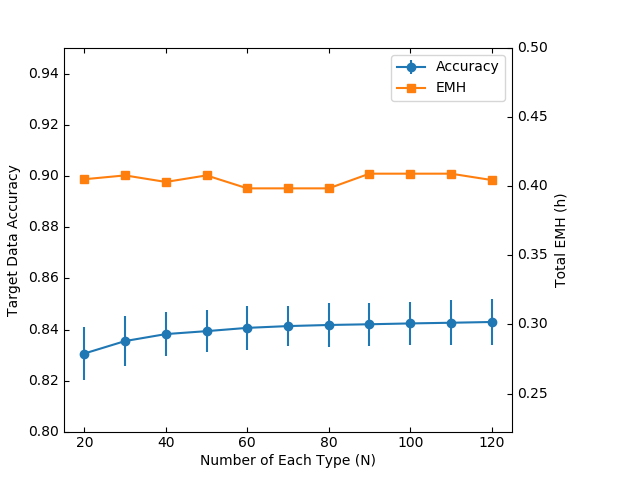}
}
\caption{SR Cost by T}
\label{fig700}
\end{figure}

The canteen scale is consistent with the product of a canteen's dish type number ($T$) and dish number of each type ($N$). The range of $T$ and $N$ are expanded in both directions simultaneously from our baseline values. The results are shown in Fig.~\ref{fig700}.

%
%

Similarities in the influence patterns of both parameters were apparent in the diagrams.
In RFID systems, both $T$ and $N$ increased with approximate linearity as their value ascended,. The speed of increase was relatively small, about $5.56\times 10^{-3}$ h/type and $2.78\times 10^{-3}$ h/dish, under low target accuracy (0.8 and lower). When the target accuracy rose beyond 0.8, however, the speed was intensified, up to $1\times 10^{-1}$ h/type and $2.78\times 10^{-2}$ h/dish at accuracy of 1.
For CV systems, both influences on the cost fluctuated only slightly, without recognizable patterns. The increment of the two features caused weak growth of average accuracy (no more than 0.02). Furthermore, the accuracy deviation reduction against boosting dish types was caused by the decreasing proportion of fixed new dish frequency($F$).

\subsubsection{New Dish Frequency}

\begin{figure}[!ht]
\subfloat[RFID Systems\label{fig80}]{%
\includegraphics[width=0.48\textwidth]{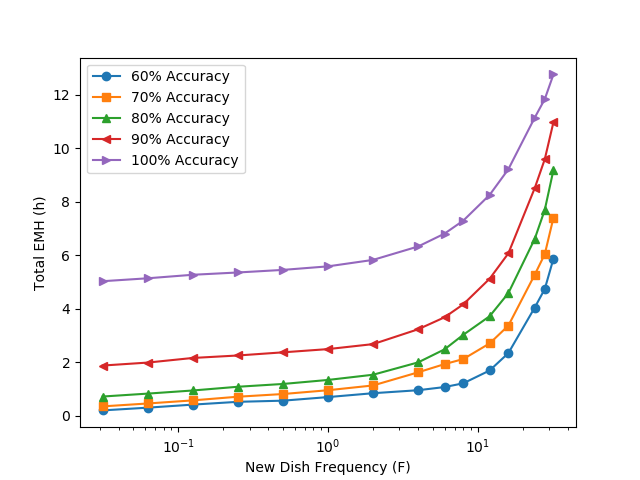}
}
\hfill
\subfloat[CV Systems\label{fig82}]{%
\includegraphics[width=0.48\textwidth]{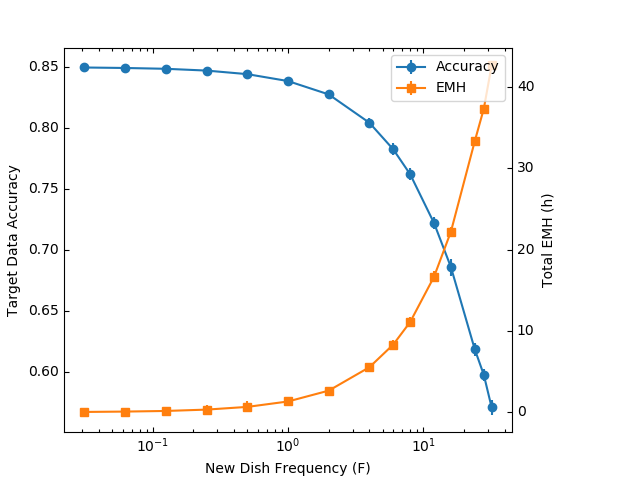}
}
\caption{SR Cost by F}
\label{fig800}
\end{figure}

The frequency of appending new dishes ($F$) can differ significantly among canteens, from only one new dish in weeks to several new dishes every meal. Therefore, we used the logarithmic scale for the F value. The results of both systems are shown in Fig.~\ref{fig80} and Fig.~\ref{fig82}.

%
The results revealed the distinct effect of $F$ on the costs of the two systems: The range of $F$ was segmented by 1 type/meal for RFID systems. The cost grew slowly in logarithm when $F$ was smaller than 1, i.e., new dishes were not added at each meal. As $F$ passed 1, the curve was computed to be approximately linear and the cost grew much faster than that of a smaller-than-one $F$. With accuracy of 1, the rate was about $2.08\times 10^{-1}$ h/type$*$meal.
However, $F$'s effect on CV systems shows uniformity, i.e., linearity along the whole range (about 1.67 h/type$*$meal). Another significant point is the reverse impact on accuracy, an approximately linear decrement with a ratio of about 0.011 /type$*$meal.

The distinction proves the far bigger influence of $F$'s value on CV systems than on RFID systems.

\subsubsection{Dish Rotation Number}

\begin{figure}[htbp]
\centerline{\includegraphics[width=2.5in]{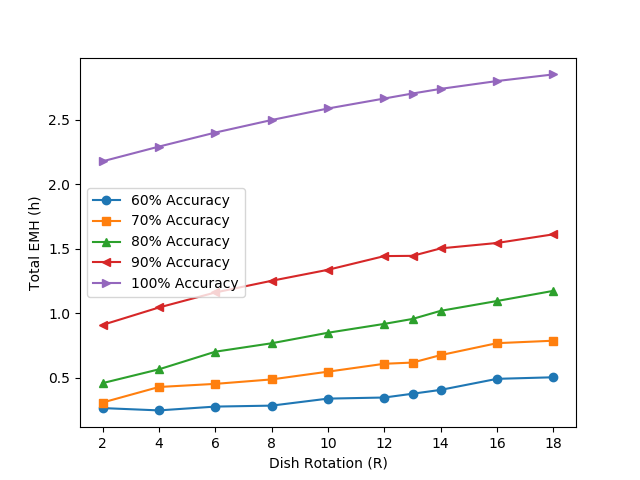}}
\caption{SR Cost by R of RFID Systems}
\label{fig90}
\end{figure}

CV systems are free of the costs of dish rotation because the features of the old dishes have been extracted and stored inside the CV model. RFID systems, however, have no way to deal with the problem except by updating the settings before or during the meal. Experiments were arranged with expanded $R$ range from the baseline, as is shown in Fig.~\ref{fig90}.

The result illustrated that with different target accuracies, the cost of RFID systems changed linearly against the value of $R$ at an approximately marginal rate, about $5.56\times 10^{-2}$ h/type$*$meal. The higher the target accuracy was, the more sparsely the curves were distributed. This proved the $R$ parameter affected the cost independently of data accuracy.

\subsubsection{Comprehensive analysis}
After four canteen features were studied, all the results were organized into Table.~\ref{tab4} for comparison.

\begin{table}
\caption{Marginal Cost (h) by Systems and Canteen Feature Effects}
\begin{center}
\begin{tabular}{r@{\quad}r@{\quad}r@{\quad}r@{\quad}rl}
\hline
\multicolumn{1}{l}{\rule{0pt}{12pt}Parameter}&\multicolumn{1}{l}{\rule{0pt}{12pt}RFID-80\%}&\multicolumn{1}{l}{\rule{0pt}{12pt}RFID-100\%}&\multicolumn{1}{l}{\rule{0pt}{12pt}CV SR Cost}&
\multicolumn{2}{l}{CV Accuracy}\\[2pt]
\hline\rule{0pt}{12pt}
T & $+1\times 10^{-1}$ & $+5.56\times 10^{-3}$ & - & $+4\times 10^{-5} \quad (T<45)$ & \\
N & $+2.78\times 10^{-2}$ & $+2.78\times 10^{-3}$ & - & $+4\times 10^{-5} \quad (N<80)$ & \\
F & $+2.08\times 10^{-1}$ & $+1.94\times 10^{-1}$ & $+1.67$ & $-1.1\times 10^{-2}$ & \\
R & $+5.56\times 10^{-2}$ & $+6.39\times 10^{-2}$ & - & - & \\
\hline
\end{tabular}
\label{tab4}
\end{center}
\end{table}

By analyzing the table, we could make the following conclusions. Firstly, it was expensive for both systems to add new dish types, but the degree was comparatively smaller for RFID systems at about one eighth of that of CV. Secondly, the cost of CV systems was directly affected by $F$, along with collateral damage to data accuracy. Finally, four features affect the cost of RFID systems to similar degrees but when target accuracy escalates, the effect from $T$ and $N$ multiplies and thus dominates.

%

\subsection{Cost in Typical Canteen Scenarios}

The values of the four features we studied have different distributions based on our dataset from real canteen situations. Four major combinations of feature values were extracted to construct four typical canteens as listed in Table.~\ref{tab5}. We noticed that the value of $N$ was relatively stable in real circumstances and the scale of the canteen was mainly indicated by $T$. Moreover, a canteen usually tended to choose either a high $F$ or $R$ to ensure menu variation and reasonable cost.
\begin{table}
\caption{Typical Canteens Based on Statistics}
\begin{center}
\begin{tabular}{r@{\quad}r@{\quad}r@{\quad}r@{\quad}r@{\quad}rl}
\hline
\multicolumn{1}{l}{\rule{0pt}{12pt}}&\multicolumn{1}{l}{\rule{0pt}{12pt}$T$}&\multicolumn{1}{l}{\rule{0pt}{12pt}$N$}&\multicolumn{1}{l}{\rule{0pt}{12pt}$F$}&\multicolumn{1}{l}{\rule{0pt}{12pt}$R$}&
\multicolumn{2}{l}{Customer number}\\[2pt]
\hline\rule{0pt}{12pt}
\textbf{TYPE I}& 20 & 70 & 0.3 & 6 & 450 & \\
\textbf{TYPE II}& 20 & 70 & 3 & 12 & 450 & \\
\textbf{TYPE III}& 50 & 70 & 0.75 & 15 & 1200 & \\
\textbf{TYPE IV}& 50 & 70 & 7.5 & 30 & 1200 & \\
\hline
\end{tabular}
\label{tab5}
\end{center}
\end{table}

\begin{figure}[!ht]
\subfloat[Type I\label{fig100}]{%
\includegraphics[width=0.24\textwidth]{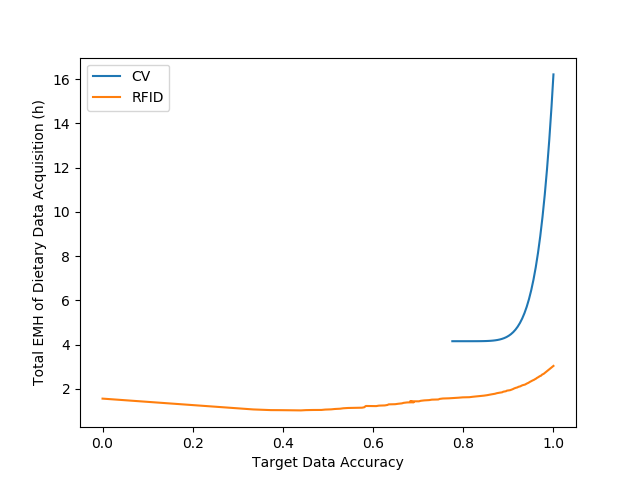}
}
\hfill
\subfloat[Type II\label{fig101}]{%
\includegraphics[width=0.24\textwidth]{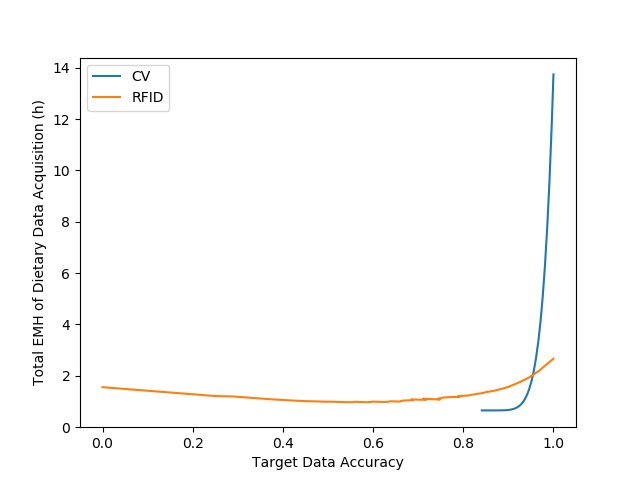}
}
\hfill
\subfloat[Type III\label{fig102}]{%
\includegraphics[width=0.24\textwidth]{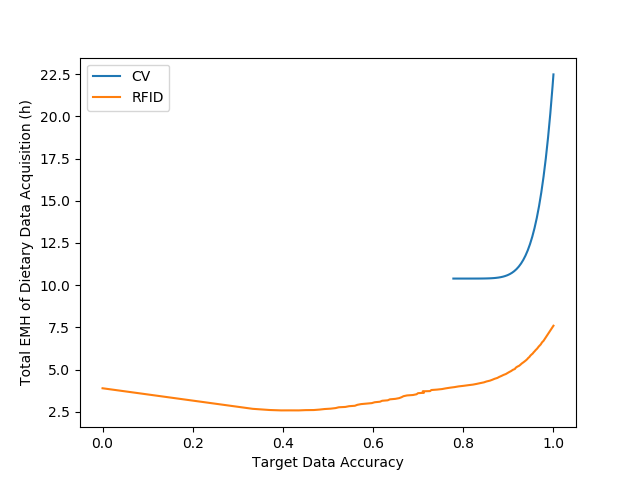}
}
\hfill
\subfloat[Type IV\label{fig103}]{%
\includegraphics[width=0.24\textwidth]{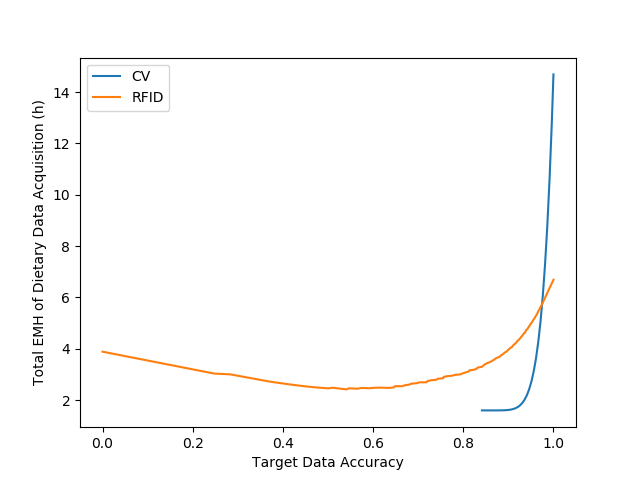}
}
\caption{Total Cost by Typical Canteens}
\label{fig10}
\end{figure}

With these four canteens, we demonstrated the cost in a more specific way. Apart from the general results explained, there were new insights as depicted in Fig.~\ref{fig10}: For a canteen with a high old dish rotation like types I and III, if data accuracy was no more than 0.95, the CV system was recommended and could save up to 50\%-75\% of costs. The bigger the canteen, the larger the cost saving. As the accuracy increased, the cost difference declined and the RFID system became better for ultimate accuracy (higher than 0.95). Meanwhile, for a canteen with a high new dish frequency, like types II and IV, the RFID system was always the better choice. Costs could be saved up to 50\% for moderate, and 70\% for large scale canteens. The cost difference decreased a little as the accuracy increased from 0.8 to 0.95.

\section{Discussions}\label{discussion}

This section discusses supplemental factors, including model parameter sensitivity, correction features and balancing, and standardization of dishes, in order to generalize our research and facilitate applications.

\subsection{Model Parameter Sensitivity: A Long Term Perspective}

The values of parameters listed in the Model section are based on our dataset and on-site measurements, which may have some deviations on a real occasion. Moreover, the change of social labor prices and technology improvements in the long term can also cause the variation of these values. To make our work more practical, extra experiments were performed to identify the extent of the influences caused by these deviations. Experiments were conducted in the baseline environment.

\subsubsection{The $S$ and $\alpha$ in EMH-A Models}\label{para-sen}

\begin{figure}[!ht]
\subfloat[SR Cost by $S$\label{fig110}]{%
\includegraphics[width=0.24\textwidth]{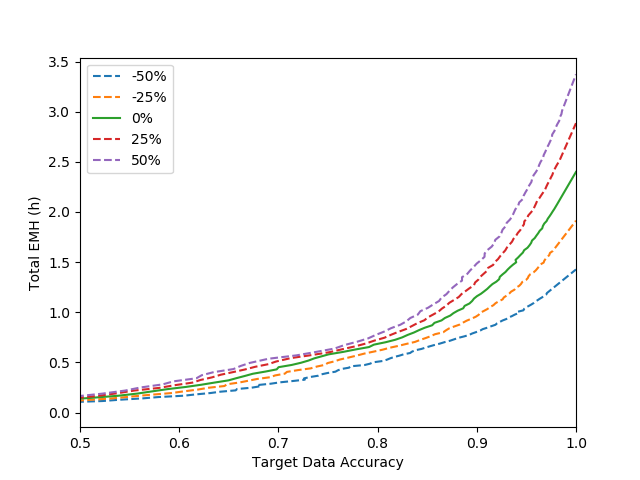}
}
\hfill
\subfloat[Unit EMH by $S$\label{fig111}]{%
\includegraphics[width=0.24\textwidth]{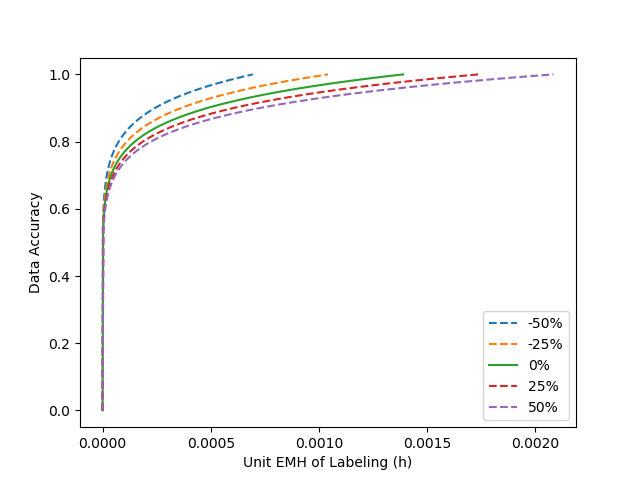}
}
\hfill
\subfloat[SR Cost by $\alpha$\label{fig112}]{%
\includegraphics[width=0.24\textwidth]{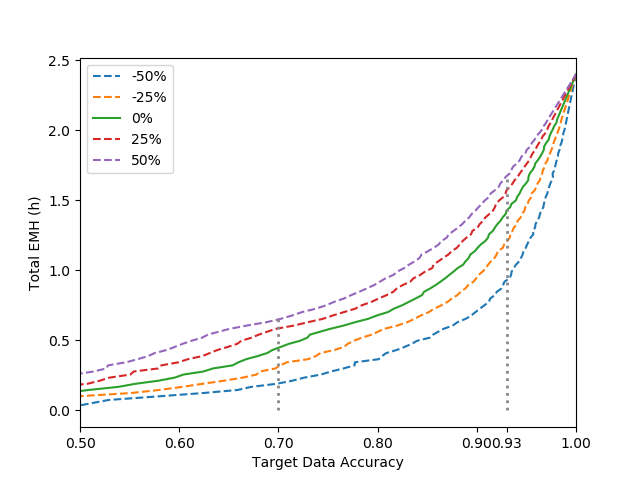}
}
\hfill
\subfloat[Unit EMH by $\alpha$\label{fig113}]{%
\includegraphics[width=0.24\textwidth]{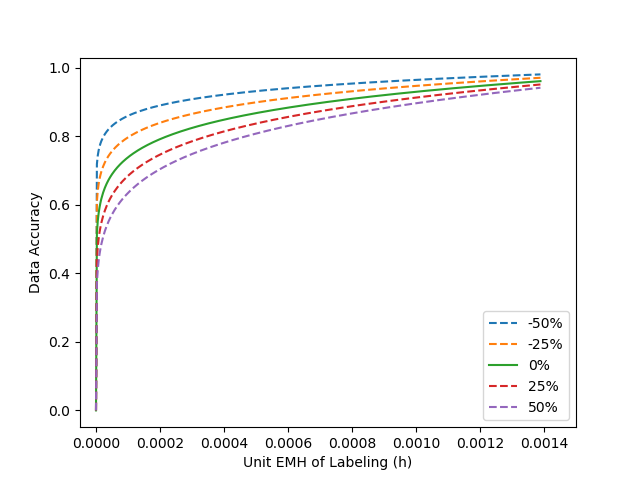}
}
\caption{SR Cost by $S$ and $\alpha$ Deviation}
\label{fig1100}
\end{figure}

The EMH-Accuracy model was most widely used throughout our research. The basic key procedures of RFID systems, namely, inputting, setting and labeling, plus correction procedures contained in both systems, all applied to this model to describe the corresponding relationship between manpower cost and data accuracy. Thus, we selected labeling, as it was found to be the dominant procedure in RFID systems, to experiment on the two parameters' value deviation effects on the total cost. The results are drawn in Fig.~\ref{fig1100}.

Further calculation based on the diagrams showed that the variation degree of total cost was proportionate to the deviation degree of $S$ by 1 accuracy, 50\% $S$ increment vs. 45\% cost increment in Fig.~\ref{fig110}. The effect decreased as the target accuracy dropped and became minor when the accuracy dropped beneath 0.8. Since the cost of labeling was dominant, the effect of $S$ deviation was smaller when it came to other procedures like inputting, setting and correcting. This also applied to  $S$ in the sampling procedure in CV systems. In terms of the $\alpha$ deviation, it mainly affected the total cost in a range of accuracy between 0.7 to 0.93, about 50\% $\alpha$ increment for 20\% total cost increment. The effect increased slowly as the $\alpha$ value decreased. In addition, the $\alpha$ decrement also led to relatively more difficulty to achieve a higher accuracy.

\subsubsection{The $\beta$ in SNA Model}

\begin{figure}[htbp]
\centerline{\includegraphics[width=0.48\textwidth]{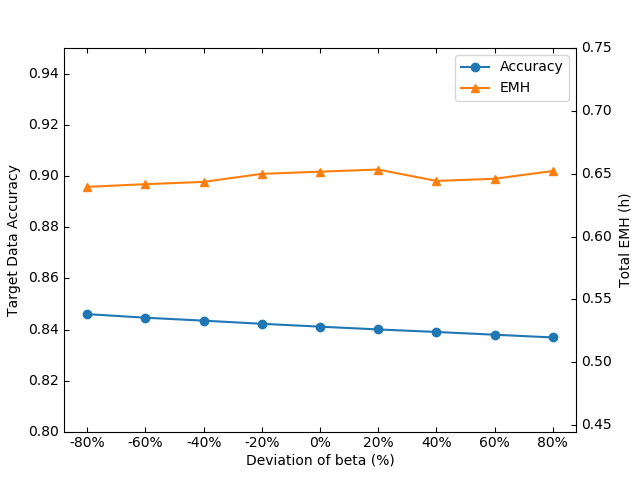}}
\caption{SP Cost and Accuracy by $\beta$ Deviation}
\label{fig120}
\end{figure}

For the SNA model in CV systems, it is predictable that experimenting on the $U$ parameter will not generate anything substantial, because of its direct bond with SP accuracy. Hence, the transmission coefficient $\beta$, also the expansion degree of our adopted sigmoid curve, was selected. Experiments were carried out in a similar way to Section~\ref{para-sen}, with results as shown in Fig.~\ref{fig120}.

The results showed that there was no obvious bond between the $\beta$ value and the total cost. However, the $\beta$ value growth caused a small accuracy decrement around 0.002 for 50\% deviation.

Finally, the long term change is also worth mentioning. As a global trend, the inevitable manpower price increases will cause $S$ to increment and thus the total cost will increase; The enhancement of systems' degree of automation will lead to a bigger $\alpha$ value which will make it more economical to pursue accurate data; The development of the CV algorithm will trigger a smaller $\beta$ and thereby higher accuracy and lower cost for CV systems.

\subsection{Correction Features: Balancing in Local Conditions}

\begin{figure}[!ht]
\subfloat[by Customer Capacity of CV Systems\label{fig130}]{%
\includegraphics[width=0.48\textwidth]{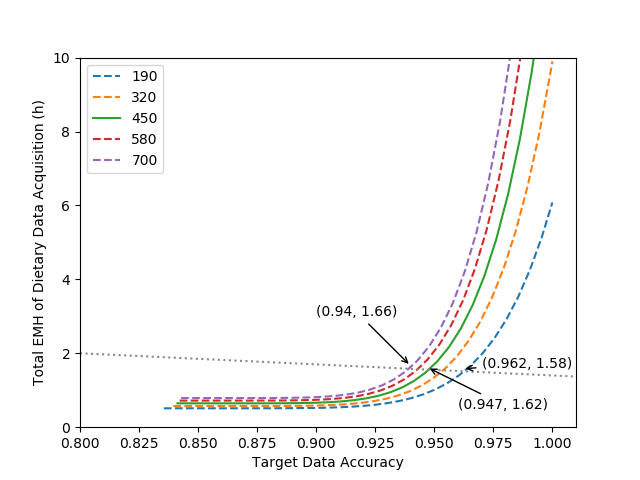}
}
\hfill
\subfloat[by Improvement Degree of RFID Systems\label{fig131}]{%
\includegraphics[width=0.48\textwidth]{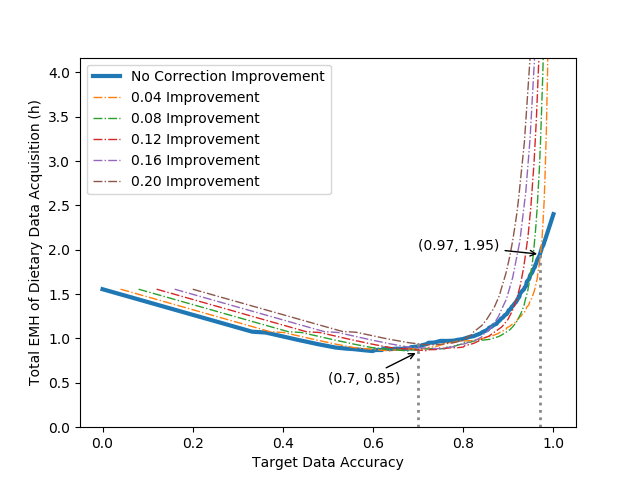}
}
\caption{Correction Cost by Customer Capacity and Improvement Degree}
\label{fig1300}
\end{figure}

As stated in Section~\ref{cv-need-crt}, data correction at checkout is the only measure for CV systems to improve their data quality. Further, correction is optional for RFID systems. The cost of correction is decided by the number of errors and expected accuracy improvement. Since CV systems have firm requirements, we experimented on canteens with different customer capacities which is proportionate with the product of $T$ and $N$, to explore its impact on correction cost. The results are depicted in Fig.~\ref{fig130}, where the dotted line is located by points with the same marginal cost growth right before the abrupt rise. Considering the efficiency of the accuracy improvement by correction, there is a dynamic limit for each canteen scale. In addition, the space for improvement lessens by about 0.022 as the canteen capacity grows from 190 to 700 customers.

For the RFID system, we constrained the accuracy improvement within 0.2 and activated the correction procedure. The results are illustrated in Fig.~\ref{fig131}. The results proved that it costs much more to depend on correction when perfect accuracy is demanded since humans are more apt to make mistakes especially in a high-pressured condition like checkout. Correction can save the cost, about 10\% maximum, when the target accuracy is between 0.7 and 0.97.

Overall, because of the elevated cost along improvement degree, correction can only bring about small cost savings in a small accuracy range and thus cannot be relied on. For CV systems, there is an efficient improvement limit which is comparatively constant. The cost of accuracy above the limit multiplies and is not economical.

%
%

\subsection{Standardization of Dish Supply: An Inevitable Trend}

\begin{figure}[htbp]
\centerline{\includegraphics[width=0.48\textwidth]{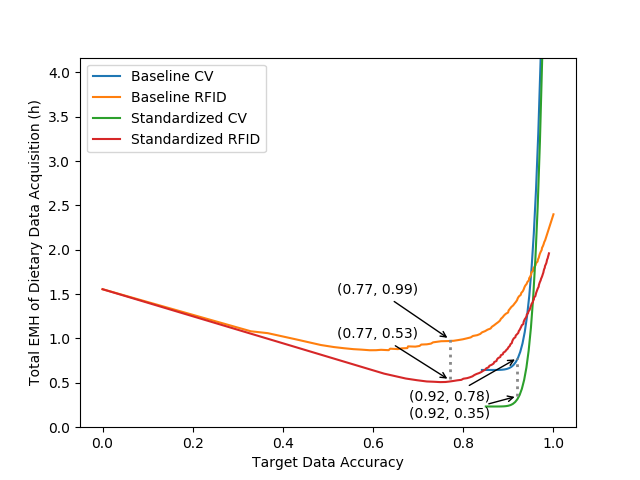}}
\caption{Cost in Standardized Canteen: Baseline canteen is used for comparison.}
\label{fig140}
\end{figure}

The frequently changed menus and recipes in modern canteens have always been the biggest liability to dietary data harvesting, and even more so for Chinese foods. Except for the extra cost paid for information inputting and on-site setting, the irregularity of the menus and the recipes also costs more in sections outside the SGC systems. Inconsistent material ordering, temporary dish interference and other expenditures can be saved if a relatively unified menu and recipes are coordinated. To determine the cost saving of dish standardization, a complementary experiment was carried out and the results are drawn in Fig.~\ref{fig140}.

So, as is shown in Fig.~\ref{fig140}, a canteen without dish addition and rotation can lead to cost savings in dietary data harvesting of up to 46.7\% for RFID systems and 55.1\% for CV systems. Against the background of emerging self-awareness of health concerns and dietary management, it seems possible for customers to compromise their preferences if effective dietary management is provided. Moreover, the dish standardization may be more economical where dietary management is of greater urgency, for example in hospitals, rehab facilities and so on. We believe that instead of passively waiting for technology developments, it would be wiser to embrace the trend of dietary management and make self-adjustments to our own habits.

\section*{Conclusions}

In this paper, we conducted an in-depth analysis of the essential mechanisms and the fundamental distinctions between RFID- and CV-based SGC systems. We analyzed the manpower costs required for dietary data acquisition, which is often overlooked. The tag binding of RFID systems leads to its long pipelined manual operations and the upper bound of CV systems' recognition models constrains system reliability, which results in unacceptable costs to maximum data accuracy. Two models based on the characteristics of staff operation in specific procedures(the EMH-A model) and sample accumulation (the SNA model) were proposed. Datasets collected from real canteens were used as input data.

In relation to the cost of dietary data acquisition, we have developed major numerical conclusions as follows:
\begin{itemize}
\item In order for accurate dietary data acquisition, a large amount of extra costs are required, of which staff operation related costs account for up to 90\%.
\item When deploying RFID systems, the cost of labeling procedures accounts for 80\% of the staff related costs. The labeling procedure is also the bottleneck to achieve perfect accuracy.
\item The accuracy of CV systems can be improved by around 0.08 to 0.92 through efficient checkout corrections, but it is unrealistic to go on forcing the accuracy to 1, since the total cost will multiply about 8 times.
\item For both types of systems, the marginal costs continued to rise when higher data accuracy was demanded. CV systems had a much higher rising rate, making its total costs bypass those of RFID systems after 0.95 accuracy.
\item It is expensive to continually introduce new dishes although RFID systems were comparatively more suitable for a moderate-sized canteen with frequent new dish additions.
\item CV systems were vulnerable to new dish additions, which increased costs by about 1.67 h / type and jeopardized the accuracy by about 0.01 / type, while there were advantages for a large-scaled.

\end{itemize}

Based on our analysis, the current advantages of RFID systems will diminish if no improvements on labeling procedures occur. CV systems can benefit from improving recognition algorithms, and higher levels of automation in sampling. They may also benefit from publicly available dietary datasets.


\section*{Acknowledgment}

We acknowledge the support of the National Natural Science Foundation of China under grant 61433009, and the National Key Research and Development Project of China under grant 2019YFC1709800.

---- Bibliography ----

\bibliographystyle{splncs03}
\bibliography{cddh}

\end{document}